\documentclass[aps,prd,twocolumn,superscriptaddress,nofootinbib,preprintnumbers]{revtex4-1}
\usepackage{amsmath}
\usepackage{graphicx}
\usepackage{subfigure}
\usepackage{amssymb}
\usepackage{xcolor}
\usepackage{multirow}
\usepackage{cancel}
\usepackage{color}
\usepackage{ulem}
\usepackage{listings}
\usepackage{xcolor}
\usepackage{float}
\usepackage{slashed}

\usepackage{tikz}
\usetikzlibrary{arrows,shapes}
\usetikzlibrary{trees}
\usetikzlibrary{matrix,arrows} 				
\usetikzlibrary{positioning}				
\usetikzlibrary{calc,through}				
\usetikzlibrary{decorations.pathreplacing}  
\usepackage{pgffor}							

\usetikzlibrary{decorations.pathmorphing}	
\usetikzlibrary{decorations.markings}
\tikzset{
    vector/.style={decorate, decoration={snake}, draw},
	provector/.style={decorate, decoration={snake,amplitude=2.5pt}, draw},
	antivector/.style={decorate, decoration={snake,amplitude=-2.5pt}, draw},
    fermion/.style={draw=black, postaction={decorate},
        decoration={markings,mark=at position .55 with {\arrow[draw=black]{>}}}},
    fermionbar/.style={draw=black, postaction={decorate},
        decoration={markings,mark=at position .55 with {\arrow[draw=black]{<}}}},
    fermionnoarrow/.style={draw=black},
    gluon/.style={decorate, draw=black,
        decoration={coil,amplitude=4pt, segment length=5pt}},
    scalar/.style={dashed,draw=black, postaction={decorate},
        decoration={markings,mark=at position .55 with {\arrow[draw=black]{>}}}},
    scalarbar/.style={dashed,draw=black, postaction={decorate},
        decoration={markings,mark=at position .55 with {\arrow[draw=black]{<}}}},
    scalarnoarrow/.style={dashed,draw=black},
    electron/.style={draw=black, postaction={decorate},
        decoration={markings,mark=at position .55 with {\arrow[draw=black]{>}}}},
	bigvector/.style={decorate, decoration={snake,amplitude=4pt}, draw},
}

\tikzstyle{block} = [draw, rectangle, 
    minimum height=3em, minimum width=6em]

\usepackage[colorlinks,citecolor=blue]{hyperref}
\usepackage{amsmath}
\usepackage{wrapfig}

\newcommand{\be}{\begin{equation}}
\newcommand{\ee}{\end{equation}}
\newcommand{\beq}{\begin{equation}}
\newcommand{\eeq}{\end{equation}}
\newcommand{\bea}{\begin{eqnarray}}
\newcommand{\eea}{\end{eqnarray}}
\newcommand{\besp}{\begin{equation}\begin{split}}
\newcommand{\eesp}{\end{split}\end{equation}}

\newcommand{\nn}{\nonumber}

\newcommand{\Eq}[1]{Eq.~(\ref{#1})}

\newcommand{\Dfbd}{\mathord{\buildrel{\lower3pt\hbox{$\scriptscriptstyle\leftrightarrow$}}\over {D}_{\mu}}}
\newcommand{\ave}[1]{\left\langle #1\right\rangle}

\def\mL{\mathcal{L}}

\def\mO{\mathcal{O}}

\def\mT{\mathcal{T}}

\def\0{\textbf{0}}
\def\1{\textbf{1}}
\def\2{\textbf{2}}
\def\3{\textbf{3}}
\def\4{\textbf{4}}
\def\5{\textbf{5}}
\def\6{\textbf{6}}
\def\7{\textbf{7}}
\def\8{\textbf{8}}
\def\9{\textbf{9}}

\def\p{\textbf{p}}

\def\hc{\text{h.c.}}

\begin{document}

\title{Primordial black holes from an electroweak phase transition}

\author{Peisi Huang,}
\email{peisi.huang@unl.edu}
\affiliation{Department of Physics and Astronomy, University of Nebraska, Lincoln, NE 68588, USA}

\author{Ke-Pan Xie}
\email{Corresponding author. kepan.xie@unl.edu}
\affiliation{Department of Physics and Astronomy, University of Nebraska, Lincoln, NE 68588, USA}

\begin{abstract}

We propose a mechanism that forms primordial black holes (PBHs) via a first-order electroweak phase transition (FOEWPT). The FOEWPT is realized by extending the Standard Model with a real singlet scalar, while the PBH formation is achieved by the collapse of non-topological solitons called Fermi-balls. Such solitons form via trapping fermions in the false vacuum during the FOEWPT, and they eventually collapse into PBHs due to the internal Yukawa attractive force. We demonstrate that a scenario with PBH dark matter candidate can exist, and the typical experimental signals include FOEWPT gravitational waves and the multi-lepton/jet or displaced vertex final states at the LHC.

\end{abstract}

\maketitle

 \section{Introduction}

Usually, ``black holes'' refer to the compact objects from the gravitational collapse of the massive stars running out of fuel; the gravity of those objects is so strong that even light cannot escape from them. However, it is proposed that black holes can also form soon after the Big Bang, well before the formation of any stars and galaxies~\cite{zel1967hypothesis,hawking1971gravitationally}. Those hypothetical black holes, known as primordial black holes (PBHs), have important cosmological implications: they could be a natural dark matter (DM) candidate~\cite{hawking1971gravitationally,Chapline:1975ojl,Khlopov:2008qy,Carr:2016drx,Carr:2020gox,Carr:2020xqk,Green:2020jor}, could be the seeds of the supermassive black holes~\cite{Bean:2002kx,Khlopov:2004sc,Duechting:2004dk,Kawasaki:2012kn,Clesse:2015wea}, or could be the origin of some gravitational wave (GW) signals observed by the LIGO/Virgo detectors~\cite{Abbott:2016blz,Abbott:2016nmj,Abbott:2017vtc,Clesse:2016vqa,Bird:2016dcv,Sasaki:2016jop}, etc.

PBHs can form in the early Universe via the collapse of overdense region from the primordial fluctuations during inflation~\cite{Carr:1974nx,carr1975primordial,Sasaki:2018dmp}, via the collapse of cosmic topological defects~\cite{Hawking:1987bn,Caldwell:1995fu,Garriga:1992nm,Rubin:2000dq,Rubin:2001yw,Dokuchaev:2004kr,Deng:2016vzb}, via scalar field fragmentation~\cite{Cotner:2016cvr,Cotner:2017tir,Cotner:2018vug,Cotner:2019ykd}, or via a first-order phase transition (FOPT)~\cite{Crawford:1982yz,Hawking:1982ga,La:1989st,Moss:1994iq,konoplich1998formation,Konoplich:1999qq,Kodama:1982sf,Lewicki:2019gmv,Kusenko:2020pcg}. Recently, there is a renewed interest in the PBH formation from an FOPT, and many mechanisms have been proposed and studied~\cite{Gross:2021qgx,Baker:2021nyl,Kawana:2021tde,Liu:2021svg,Davoudiasl:2021olb,Baker:2021sno,Jung:2021mku,Hashino:2021qoq,Marfatia:2021hcp,Maeso:2021xvl}. Especially, Ref.~\cite{Kawana:2021tde} proposes a general mechanism that non-topological solitons called ``Fermi-balls'' form during an FOPT, and then collapse into PBHs due to the internal Yukawa attractive force. In this article, we would like to apply this mechanism to the extended Standard Model (SM), discussing the possibility of forming PBHs in a first-order electroweak phase transition (FOEWPT).

\begin{figure*}
\centering
\includegraphics[scale=0.42]{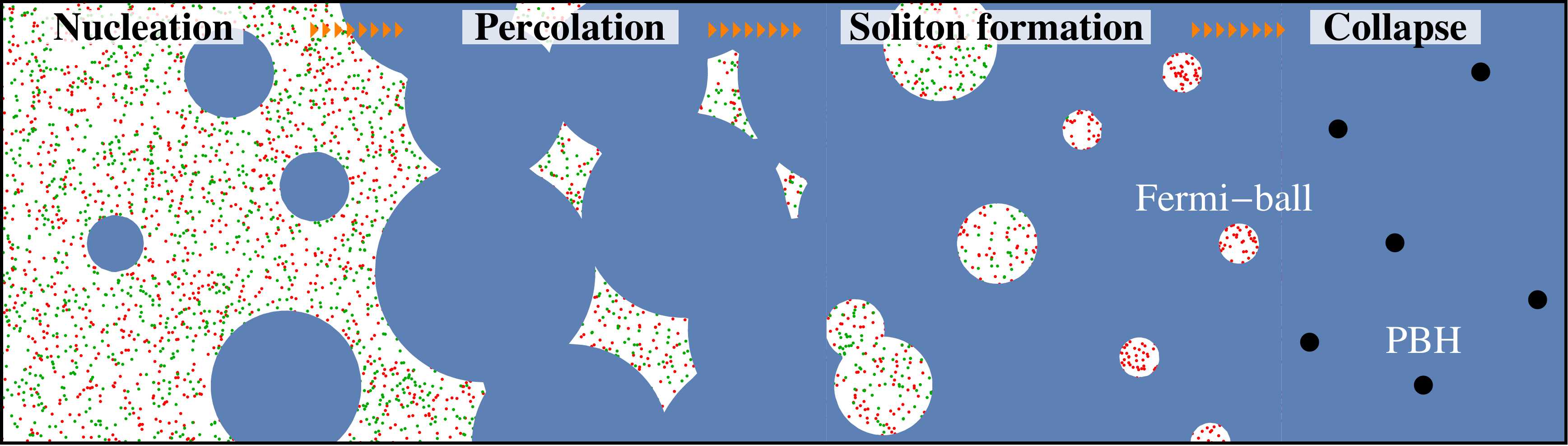}
\caption{Sketch of the mechanism, where white and blue color regions represent the false and true vacua, and red and green dots represent $\chi$ and $\bar\chi$, respectively. The FOEWPT proceeds by bubble nucleation and percolation. Soon after percolation, the trapped fermions are squeezed into small false vacuum remnants to form Fermi-balls. After the completion of FOEWPT, the Fermi-balls cool down and collapse into PBHs.}
\label{fig:sketch}
\end{figure*}

The SM EWPT is a smooth crossover~\cite{Kajantie:1996qd,Rummukainen:1998as,Laine:1998jb}. However, an FOEWPT can be realized if the SM is simply extended with a real singlet scalar $S$~\cite{McDonald:1993ey,Espinosa:2011ax,Cline:2012hg,Alanne:2014bra,Cheng:2018ajh,Carena:2019une}. If we further extend this model with one fermion $\chi$ coupling to the scalar via $-y_\chi S\bar\chi\chi$, then $\chi$ would have different masses inside and outside the vacuum bubbles during the FOEWPT, as the $S$ vacuum expectation values (VEVs) are different in two sides of the bubble wall. This mass gap, if significantly larger than the FOEWPT temperature, would forbid the $\chi$ fermions from penetrating into the true vacuum (i.e. EW symmetry breaking phase). After the completion of the FOEWPT, the fermions are trapped in the false vacuum and then form non-topological solitons, dubbed Fermi-balls, if there is a number density asymmetry for $\chi$ and $\bar\chi$~\cite{Hong:2020est}. Inside a Fermi-ball, the constituent $\chi$'s interact with each other via the $S$-mediated attractive Yukawa force, and the corresponding range of force increases as the Fermi-ball cools down. When the range of force reaches the mean separation of $\chi$'s in a Fermi-ball, the ball collapses into a PBH~\cite{Kawana:2021tde}. The mechanism is sketched in Fig.~\ref{fig:sketch}.

This article is not just a simple application of an existing mechanism. The original study~\cite{Kawana:2021tde} illustrates the mechanism with a toy model with a single-field induced FOPT, adopts the preexisting $\chi$-asymmetry as an assumption, and demonstrates that the PBHs are typically overproduced compared to the DM relic abundance. In this work, we demonstrate the nontrivial features caused by a two-field induced FOPT, and build a concrete model which can generate the $\chi$-asymmetry and provide necessary dilution process to realize a PBH DM scenario. We will first discuss the FOEWPT dynamics, Fermi-ball and PBH formation in Section~\ref{sec:FOPT_PBH}, and then build the complete model in Section~\ref{sec:UV}. After discussing the phenomenology of the model in Section~\ref{sec:pheno}, we conclude in Section~\ref{sec:conclusion}.

\section{The FOEWPT and PBH formation}\label{sec:FOPT_PBH}

\subsection{The FOEWPT dynamics}\label{sec:FOPT}

Denote the Higgs doublet as $H=(\sqrt2G^+,h+iG^0)^T/\sqrt{2}$ and the real singlet as $S$, the scalar sector of the model reads
\be\label{L}
\mL\supset D_\mu H^\dagger D^\mu H+\frac12\partial_\mu S\partial^\mu S-V(H,S),
\ee
where the joint scalar potential is
\begin{multline}\label{V}
V(H,S)=-\mu^2|H|^2+\lambda|H|^4+\frac{a_1}{2}|H|^2S\\
+\frac{a_2}{2}|H|^2S^2+\frac{b_2}{2}S^2+\frac{b_3}{3}S^3+\frac{b_4}{4}S^4.
\end{multline}
We have shifted $S$ such that the tadpole term $b_1S$ vanishes. At $T=0$, the above potential has a VEV $(h,S)=(v,v_s)$, where $v=246$ GeV. Shifting $h\to v+h$ and $S\to v_s+S$, one gets the mass term of $h$ and $S$ via the Hessian matrix of the potential. Diagonalizing the mass term yields two mass eigenstates,
\be
\begin{pmatrix}h\\ S\end{pmatrix}=\begin{pmatrix}\cos\theta&-\sin\theta\\ \sin\theta&\cos\theta\end{pmatrix}\begin{pmatrix}h_1\\ h_2\end{pmatrix}.
\ee
Here we define $h_1$ to be the Higgs-like boson discovered at the LHC~\cite{ATLAS:2012yve,CMS:2012qbp}, thus $M_{h_1}=125.09$ GeV and the mixing angle $\theta$ is expected to be small. Given $M_{h_1}$ and $v$, there are 5 free parameters in \Eq{V}. We use the strategy the same as Ref.~\cite{Liu:2021jyc} to find the parameter space satisfying the SM measurements (i.e. $M_{h_1}$ and $v$). When scanning, we keep $M_{h_2}\in[250,1000]~{\rm GeV}$ and $\theta\in[0,0.35]$, and the other 3 potential parameters are within the unitarity bound and bounded-below range.

At finite temperature, the potential \Eq{V} is modified to
\begin{multline}\label{VT}
V(h,S,T)=-\frac{\mu^2-c_HT^2}{2}h^2+\frac{\lambda}{4}h^4+\frac{a_1}{4}h^2S+\frac{a_2}{4}h^2S^2\\
+m_1T^2S+\frac{b_2+c_ST^2}{2}S^2+\frac{b_3}{3}S^3+\frac{b_4}{4}S^4,
\end{multline}
under the unitary gauge, where only the gauge invariant $T^2$-order terms are kept~\cite{Dolan:1973qd,Braaten:1989kk}, and the coefficients
\be\begin{split}
c_H=&~\frac{3g^2+g'^2}{16}+\frac{y_t^2}{4}+\frac{\lambda}{2}+\frac{a_2}{24},\\
c_S=&~\frac{a_2}{6}+\frac{b_4}{4},\quad
m_1=\frac{a_1+b_3}{12}.
\end{split}\ee
For appropriate parameter choice, the thermal potential \Eq{VT} is able to trigger an FOEWPT from the false vacuum $(h,S)=(0,v_s^i)$ to the true vacuum $(v^f,v_s^f)$.

An FOEWPT is the decay between two vacua separated by a barrier. The Universe is initially in the EW symmetry preserving vacuum $(0,v_s^i)$. Below the critical temperature $T_c$, the EW symmetry breaking vacuum $(v^f,v_s^f)$ has a lower energy, thus the system acquires a decay probability per unit volume
\be\label{Gamma}
\Gamma(T)\sim T^4e^{-S_3(T)/T},
\ee
where $S_3(T)$ is the Euclidean action of the $O(3)$-symmetric bounce solution~\cite{Linde:1981zj}. The FOEWPT proceeds via bubble nucleation and percolation, where nucleation happens when the transition probability in a Hubble volume and a Hubble time reaches $\mO(1)$, i.e. $\Gamma(T_n)H^{-4}(T_n)\approx1$, while percolation happens at $T_p$ when the volume fraction of the false vacuum falls to $p_p=0.71$ that the connected bubbles are able to form an infinite cluster~\cite{rintoul1997precise}. As we will see, the phase transition considered in this article is not an ultra-supercooling one, thus the nucleation and percolation temperatures are quite close that we treat them as the same value, $T_n\approx T_p$. The Hubble constant $H(T)$ is given by $2\pi\sqrt{\pi g_*/45}(T^2/M_{\rm Pl})$ in the radiation domination era, where $g_*=106.75$ is the number of relativistic degrees of freedom.

\begin{figure}
\centering
\includegraphics[scale=0.25]{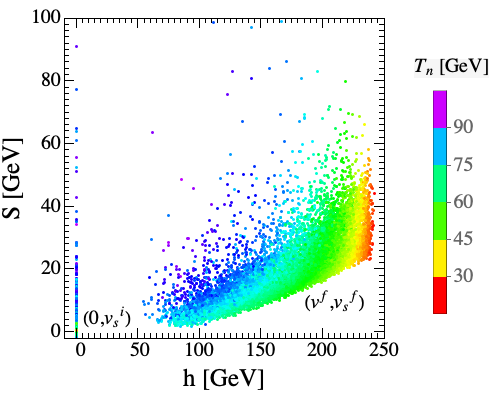}
\includegraphics[scale=0.235]{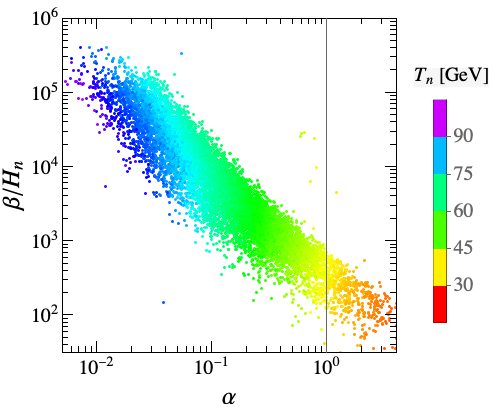}
\caption{Left: the collection of FOEWPT initial vacuum $(0,v_s^i)$ and final vacuum $(v_f,v_s^f)$. Right: the collection of $\alpha$ and $\beta/H_n$ parameters. The transition temperature $T_n$ is shown in color.}
\label{fig:FOEWPT}
\end{figure}

The potentials satisfying SM constraints are fed to the {\tt CosmoTransitions} package~\cite{Wainwright:2011kj} to calculate $S_3(T)$, and the nucleation temperature is determined by~\cite{Quiros:1999jp}
\be
\frac{S_3(T)}{T}\Big|_{T_n}\approx140.
\ee
Above equation is defined the criterion of FOEWPT. The parameter space we found for FOEWPT is demonstrated in the left panel of Fig.~\ref{fig:FOEWPT} by the scatter plot of the initial (false) vacuum $(0,v_s^i)$ and new (true) vacuum $(v_f,v_s^f)$. $T_n$ is shown in color. In the right panel of Fig.~\ref{fig:FOEWPT}, we show two important parameters of the transition, namely
\be\begin{split}\label{alphabeta}
\alpha=&~\frac{1}{g_*\pi^2T_n^4/30}\left(T\frac{\partial U_0}{\partial T}-U_0\right)\Big|_{T_n};\\
\frac{\beta}{H_n}=&~T\frac{d(S_{3}/T)}{dT}\Big|_{T_n},
\end{split}\ee
where $U_0(T)=V(0,v_s^i,T)-V(v_f,v_s^f,T)$ is the positive free energy difference between the true and false vacua. By definition, $\alpha$ and $\beta/H_n$ are ratios of FOEWPT latent heat to radiation energy density and Hubble time scale to FOEWPT duration, respectively. After the FOEWPT, as $T$ falls, the vacuum eventually shifts to current $T=0$ value $(v,v_s)$.

\subsection{Fermi-ball formation during the FOEWPT}\label{sec:FB}

Consider the fermion sector. Let $\chi$ be the singlet fermion, the relevant Lagrangian is
\be\label{L_fermi}
\mL\supset\bar\chi(i\slashed{\partial}-M_0)\chi-y_\chi S\bar\chi\chi,
\ee
thus during the FOEWPT the fermion masses in the false and true vacua are
\be\label{f_mass}
M_i=|M_0+y_\chi v_s^i|,\quad M_f=|M_0+y_\chi v_s^f|,
\ee
respectively. If $M_f-M_i\gg T_n$, then the fermions do not have sufficient kinetic energy to pass the bubble wall to enter the true vacuum. Instead, they are trapped in the false vacuum. The trapping fraction $F_\chi^{\rm trap}$ can be derived as a function of $(M_i,M_f,T_n,v_b,v_+)$, where $v_b$ and $v_+$ are the wall velocities relative to the plasma \emph{at infinite distance} and \emph{just in front of the wall}, respectively; in general $v_+\leqslant v_b$~\cite{No:2011fi}. The detailed calculation is given in Appendix~\ref{app:trapping}. For mass gap over temperature ratio $(M_f-M_i)/T_n\sim \mO(10)$, the trapping is very efficient. For example, if $M_i=0$, $M_f/T_n=10$, $v_b=0.4$ and $v_+=0.2$, then $F_\chi^{\rm trap}=99.8\%$.

As the true vacuum bubbles expand and merge, occupying more and more space, the false vacuum remnants are separated into individual pockets. This happens at $T_*$ when the volume fraction of the false vacuum decreases to $p_*=0.29$~\cite{Hong:2020est}. For a non-ultra-supercooling transition, $T_*$ is usually very close to $T_n$ and $T_p$, and we use $T_*\approx T_n$ throughout this paper. Those separated false vacuum pockets first split to smaller ones, then shrink to a negligible size. During such shrinking, the trapped fermions are forced to annihilate via $\chi\bar\chi\to S$ and $\chi\bar\chi\to SS$, while the $S$'s eventually annihilate/decay to SM particles. If there is a preexisting $\chi$-$\bar\chi$ number density asymmetry, then $\chi$'s can survive the annihilation and develop a degeneracy pressure. Once such pressure is able to balance the vacuum pressure, the Fermi-balls form~\cite{Hong:2020est}.\footnote{It is worth mentioning the difference between our work and Ref.~\cite{Baker:2021nyl}. While both considering trapping fermions during an FOPT using the large mass gap, Ref.~\cite{Baker:2021nyl} assumes either a tiny Yukawa ($y_\chi=10^{-5}\sqrt{T_n/{\rm PeV}}$) or a high temperature ($T_n=10^{15}$ GeV) to suppress the annihilation cross sections $\sigma(\chi\bar\chi\to S)$ and $\sigma(\chi\bar\chi\to SS)$, such that the $\chi$ and $\bar\chi$ number in a false vacuum remnant does not change, and hence during shrinking the remnant's energy density increases rapidly to cause the direct collapse into a PBH. In contrast, we use $y_\chi\sim\mO(1)$ and $T_n\sim100$ GeV, so that the $\chi\bar\chi$ annihilation is very efficient and no significance overdensity is formed during the shrinking, and the shrinking stops only when the surviving $\chi$'s develop sufficient degeneracy pressure to form the Fermi-ball.} Below we perform quantitative calculation for above physical picture.

First, we consider a false vacuum pocket at the end of splitting and the beginning of shrinking. The radius $R_*$ of such a pocket is determined by the consideration that {\it it should shrink to a negligible size before another true vacuum bubble is created inside it}, i.e.
\be\label{R*}
\Gamma(T_n)\left(\frac{4\pi}{3}R_*^3\right)\frac{R_*}{v_b}\sim1,
\ee
from which we can also infer the number density of those pockets $n_{\rm rem}^*=p_*/(4\pi R_*^3/3)$. Since one such pocket shrinks to one Fermi-ball, $n_{\rm rem}^*$ is also the Fermi-ball number density at formation $n_{\rm FB}^*$.

Second, we turn to the fermion number trapped in a Fermi-ball. Define the $\chi$-asymmetry in a way similar to the SM baryon asymmetry as
\be
\eta_\chi=\frac{n_\chi-n_{\bar\chi}}{s},
\ee
with $s(T)\equiv(2\pi^2/45)g_*T^3$ being the entropy density, then after annihilation there are
\be\begin{split}
Q_{\rm FB}=&~F_\chi^{\rm trap}\frac{\eta_\chi s(T_n)}{p_*}\left(\frac{4\pi}{3}R_*^3\right)\\
=&~F_\chi^{\rm trap}\frac{\eta_\chi s(T_n)}{p_*}\left(\frac{4\pi}{3}\right)^{1/4}\left(\frac{v_b}{\Gamma(T_n)}\right)^{3/4},
\end{split}\ee
$\chi$'s survive in a Fermi-ball, and all $\bar\chi$'s are gone. $Q_{\rm FB}$ is also the net $Q$-charge collected by a Fermi-ball, as the Lagrangian~(\ref{L_fermi}) has a $U(1)_Q$ invariance for $\chi\to\chi e^{i\alpha}$.

Finally, we obtain the Fermi-ball profile by solving the balance between the Fermi-gas pressure and vacuum pressure. This can be done by deriving the Fermi-ball energy $E_{\rm FB}$ under a given charge $Q_{\rm FB}$, radius $R$ and temperature $T$, and then varying $R$ to find the balance point $dE_{\rm FB}/dR=0$. For a grand canonical ensemble that consists of non-interacting fermions with desperation relation $\epsilon=\sqrt{|\p|^2+M_i^2}$, the grand potential density is
\be\label{grand_potential}\begin{split}
\tilde\omega=&~-2T\int\frac{d^3p}{(2\pi)^3}\ln\left(1+e^{-(\epsilon-\mu)/T}\right)\\
=&~-\frac{1}{3\pi^2}\int_{M_i}^\infty\frac{(\epsilon^2-M_i^2)^{3/2}d\epsilon}{e^{(\epsilon-\mu)/T}+1}.
\end{split}\ee
Assuming the chemical potential $\mu\gg T$ (which is reasonable because there are only $\chi$'s left inside the Fermi-ball), the above expression can be calculated via the low temperature expansion of the Fermi integral to be
\begin{multline}
\tilde\omega\approx-\frac{1}{24\pi^2}\left[\mu \sqrt{\mu^2-M_i^2} (2\mu^2-5 M_i^2)\right. \\
\left.+3M_i^4 \text{\,arccosh\,}\frac{\mu}{M_i}\right]-\frac{1}{6}T^2\mu\sqrt{\mu^2-M_i^2}.
\end{multline}
Using the grand potential, one is able to calculate other observables of the system, e.g. the total fermion number $Q_{\rm FB}$, the fermion energy $E_{\rm kin}$, etc. Given the relation between $\mu$ and $Q_{\rm FB}$, we can use $Q_{\rm FB}$ to rewrite the kinetic energy as
\begin{multline}\label{Ekin}
E_{\rm kin}=\frac{3\pi}{4}\left(\frac{3}{2\pi}\right)^{2/3}\frac{Q_{\rm FB}^{4/3}}{R}\left[\sqrt{1+4\delta^2}\left(1+2\delta^2\right)\right.\\
\left.-\frac{6}{\pi}\left(\frac{3}{2\pi}\right)^{1/3}\delta^4\text{\,arccsch\,}\left(2\delta\right)\right.\\
\left.+\frac{8\pi^2}{3}\delta^2\sqrt{1+4\delta^2}\left(\frac{T}{M_i}\right)^2\right],
\end{multline}
where $\delta\equiv(M_iR)/(18\pi Q_{\rm FB})^{1/3}$ is expected to be small. It is clear in \Eq{Ekin} that the fermion energy consists of the Fermi-gas kinetic part (irrelevant to $T$) and the thermal excitation part (proportional to $T^2$).

The total energy of a Fermi-ball is
\be\label{EFB}
E_{\rm FB}=E_{\rm kin}+4\pi\sigma_0 R^2+\frac{4\pi}{3}U_0R^3,
\ee
where the second term is the negligible surface tension term (because it turns out that a Fermi-ball has a macroscopic size), while the third term is the bulk energy defined below \Eq{alphabeta}. The physical radius and mass of the Fermi-ball can be determined by 
\be\label{E_stable}
\frac{dE_{\rm FB}}{dR}\Big|_{R_{\rm FB}}=0,\quad M_{\rm FB}=E_{\rm FB}\big|_{R_{\rm FB}},
\ee
but as \Eq{Ekin} involves the non-polynomial functions for $R$, analytical expressions can be got only under the small $\delta$ expansion up to $\mO(\delta^2)$, which yields
\begin{multline}
E_{\rm kin}\approx\frac{3\pi}{4}\left(\frac{3}{2\pi}\right)^{2/3}\frac{Q_{\rm FB}^{4/3}}{R}\left[\left(1+4\delta^2\right)\right.\\
\left.+\frac{8\pi^2}{3}\delta^2\left(1+2\delta^2\right)\left(\frac{T}{M_i}\right)^2\right],
\end{multline}
and hence the Fermi-ball radius can be resolved analytically
\begin{widetext}
\bea\label{MR_T_profile}
R_{\rm FB}&\approx&Q_{\rm FB}^{1/3}\left[\frac{3}{16}\left(\frac{3}{2\pi}\right)^{2/3}\frac{1}{U_0}\right]^{1/4}\left(1-\frac{M_i^2}{8\sqrt{3}\pi U_0^{1/2}}\right)\left[1-\left(\frac{\pi}{12\sqrt{3}\,U_0^{1/2}}+\frac{M_i^2}{48U_0}\right)T^2\right],\nn\\
M_{\rm FB}&\approx&Q_{\rm FB}\left(12\pi^2U_0\right)^{1/4}\left(1+\frac{\sqrt{3}M_i^2}{8\pi U_0^{1/2}}\right)\left[1+\left(\frac{\pi}{4\sqrt{3}\,U_0^{1/2}}-\frac{M_i^2}{48U_0}\right)T^2\right].
\eea
\end{widetext}
When taking $M_i\to0$ and $T\to0$, above profiles reduce to those in Ref.~\cite{Hong:2020est}.

Equation~(\ref{E_stable}) only ensures the Fermi-ball's stability under the variation of radius. To be a really stable soliton, the profile should further satisfy
\be\label{stability}
\frac{dM_{\rm FB}}{dQ_{\rm FB}}<M_f,\quad \frac{d^2M_{\rm FB}}{dQ_{\rm FB}^2}<0,
\ee
so that the Fermi-ball can be stable against decay and fission. The first condition needs to be verified for a concrete model but the second one is automatically satisfied once the surface tension ($\propto Q_{\rm FB}^{2/3}$) is taken into account. After formation, Fermi-ball can cool down via the emission of light SM fermions, and the cooling time scale is much shorter than the Hubble time scale~\cite{Kawana:2021tde}. As a result, Fermi-balls can track the cosmic temperature, and hence the mass and radius profiles change slowly with the temperature according to \Eq{MR_T_profile}.

\begin{figure}
\centering
\includegraphics[scale=0.305]{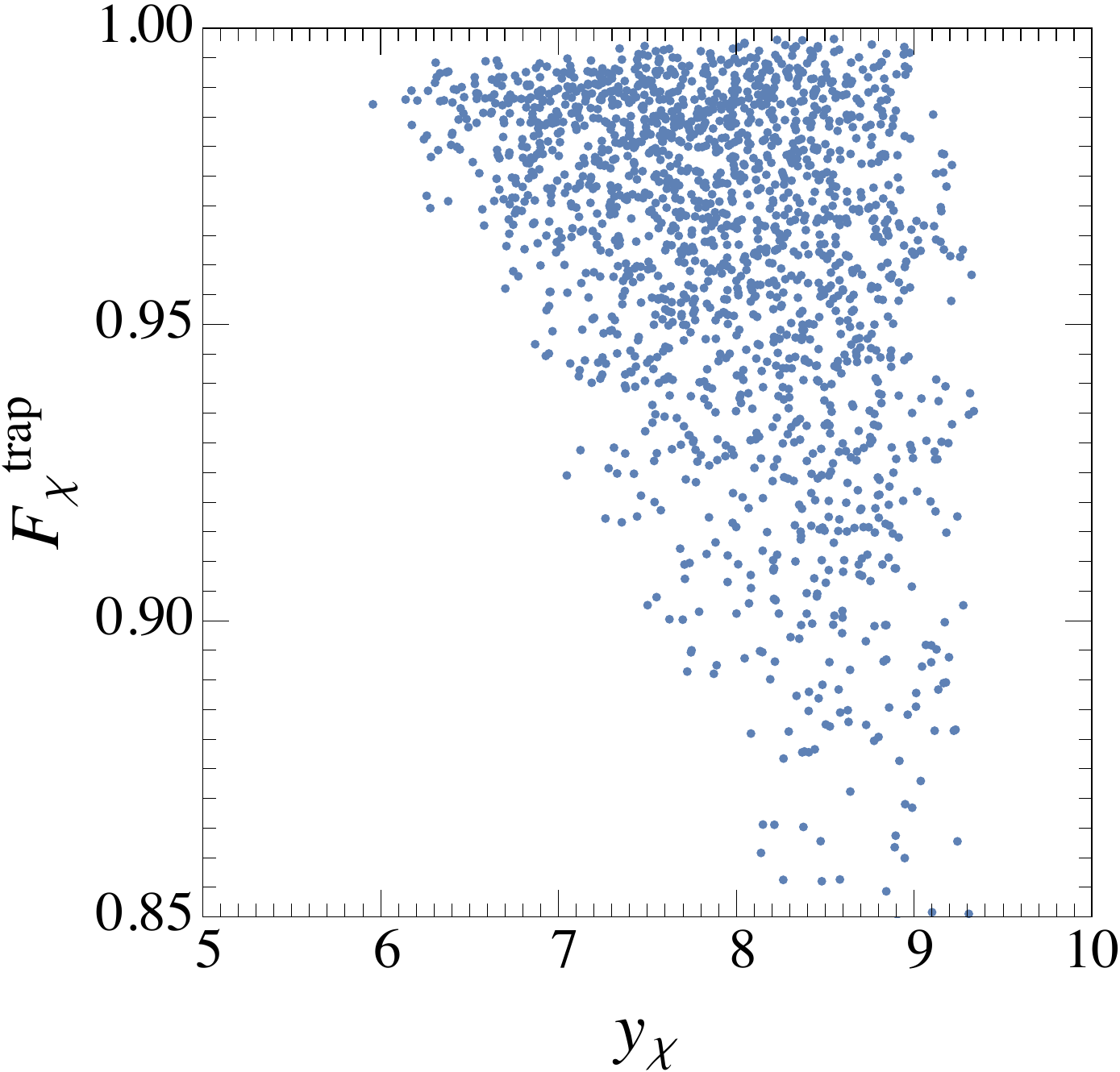}
\includegraphics[scale=0.29]{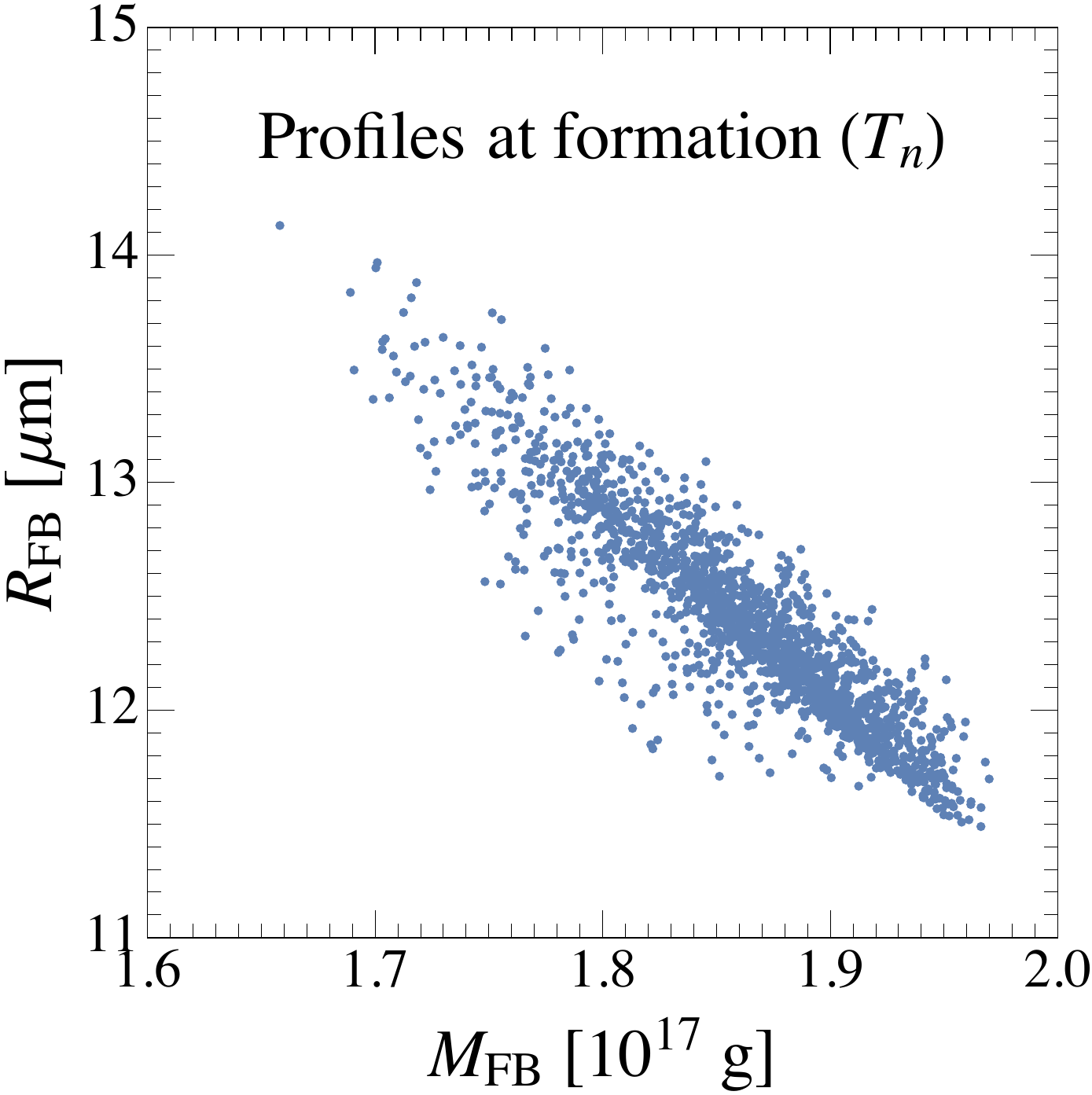}
\caption{Left: the Yukawa coupling and trapping fraction of the Fermi-ball formation data points. Right: the collection of Fermi-ball profiles at formation temperature $T_n$.}
\label{fig:FB}
\end{figure}

Now we investigate the possibility of forming Fermi-balls in the FOEWPT data points derived in Section~\ref{sec:FOPT}. For simplicity, we set the bare fermion mass $M_0=0$. Note that this choice of $M_0$ does not mean the mass in false vacuum $M_i=0$, because $\ave{S}$ is generally nonzero in the false vacuum, see \Eq{f_mass} and Fig.~\ref{fig:FOEWPT}. The wall velocity $v_b$ is important in this aspect, as it affects both the trapping fraction and the Fermi-ball mass, i.e. $M_{\rm FB}\propto v_b^3$~\cite{Kawana:2021tde}. As Ref.~\cite{Kawana:2021tde} shows that $v_b$ typically varies from 0.2 to 0.8 for a phase transition at EW scale, we adopt $v_b=0.4$ as a benchmark, and derive $v_+$ by solving the hydrodynamics profile~\cite{Espinosa:2010hh}. For each set of FOEWPT parameters, we randomly assign a $y_\chi\in[2,4\pi]$ and calculate $F_\chi^{\rm trap}$. The $\chi$-asymmetry is chosen ad $\eta_\chi=10^{-8}$. With above values in hand, together with the decay rate $\Gamma(T_n)$, one is able to derive the Fermi-ball profile~(\ref{MR_T_profile}) and check the stability conditions (\ref{stability}). In Fig.~\ref{fig:FB} we present the data points allowing the Fermi-ball formation at $T_n$. We can see that a fairly large $y_\chi\sim8$ is needed for efficient trapping,\footnote{Such a large $y_\chi$ may cause the Landau pole problem at TeV scale, and we will comment on this at the end of the conclusion.} and the mass and radius of Fermi-balls are $M_{\rm FB}\sim10^{17}$ g and $R_{\rm FB}\sim10~\mu$m, respectively. The charge of a single Fermi-ball is $Q_{\rm FB}\sim10^{38}$. The profiles shown in the right panel of Fig.~\ref{fig:FB} are evaluated at $T_n$. When $T$ falls, the profiles change mildly, until the possible collapse to PBHs, as discussed in the following subsection.

\subsection{From Fermi-balls to PBHs}\label{sec:PBH}

Section~\ref{sec:FB} has assumed the $\chi$ fermions inside a Fermi-ball are independent particles so that the Fermi-Dirac statistics applies. However, in principle the $\chi$ fermions attract each other via the $S$-mediated Yukawa force due to the interaction $-y_\chi S\bar\chi\chi$. More precisely, after the $h$-$S$ mixing there are two Yukawa forces given by the potential
\be
V_{\rm Yuk}(r)=-\frac{y_\chi^2s_\theta^2}{4\pi r}e^{-M_{h_1}r}-\frac{y_\chi^2c_\theta^2}{4\pi r}e^{-M_{h_2}r},
\ee
with $r$ being the distance between two $\chi$ fermions, and $c_\theta$ ($s_\theta$) short for $\cos\theta$ ($\sin\theta$). The Yukawa potential decays quickly when $r$ is larger than the range of force, which equals to the inverse of the mediator mass. This subsection only analyzes the parameter points in which the Yukawa interactions are negligible during the Fermi-ball formation, so that the calculations in Section~\ref{sec:FB} are valid.\footnote{The case that Yukawa interactions are too strong to form solitons is discussed in the conclusion.} However, as we will see, $M_{h_{1,2}}$ decreases when the Fermi-ball cools down, and hence the range of Yukawa force increases. When the internal Yukawa force is strong enough, a Fermi-ball cannot maintain its stability; instead, it will collapse to a PBH, somewhat similar to the stellar collapse due to gravity.

Let us quantitatively derive the collapse condition for a Fermi-ball. Using the uniform distribution of $\chi$ as a zero-order approximation, the Yukawa energy of a Fermi-ball is~\cite{Kawana:2021tde}
\be\label{EYuk}
E_{\rm Yuk}=-\frac{3y_\chi^2}{20\pi}\frac{Q_{\rm FB}^2}{R_{\rm FB}}\left(s_\theta^2f\left(\frac{L_{h_1}}{R_{\rm FB}}\right)+c_\theta^2f\left(\frac{L_{h_2}}{R_{\rm FB}}\right)\right),
\ee
where
\be
f(\xi)=\frac{5}{2}\xi^2\left[1+\frac{3}{2}\xi\left(\xi^2-1\right)-\frac{3}{2}\xi(\xi +1)^2e^{-2/\xi}\right],
\ee
satisfying $f(0)=0$ and $f(\infty)=1$. Here $L_{h_i}=1/M_{h_i}$ with $i=1$, 2 is the range of Yukawa force. We can see $E_{\rm Yuk}<0$, since the Yukawa interaction is attractive; and $E_{\rm Yuk}$ vanishes in the limit $L_{h_i}\to0$. But on the other hand, $E_{\rm Yuk}$ is enhanced by $Q_{\rm FB}^2$, which has a higher power dependence on $Q_{\rm FB}$ compared to the Fermi-gas kinetic energy \Eq{Ekin}. As a Fermi-ball collects a huge amount of $Q_{\rm FB}$, \Eq{EYuk} might dominate the total energy, causing the collapse into a PBH.

Let's consider the case of $L_{h_{1,2}}\ll R_{\rm FB}$, and then \Eq{EYuk} can be approximated as
\be\label{EYuk_sim}
E_{\rm Yuk}\approx-\frac{3y_\chi^2}{8\pi}\frac{Q_{\rm FB}^2}{R_{\rm FB}^3}\left(\frac{s_\theta^2}{M_{h_1}^2}+\frac{c_\theta^2}{M_{h_2}^2}\right)\equiv-\frac{3y_\chi^2}{8\pi}\frac{Q_{\rm FB}^2}{R_{\rm FB}^3}\frac{1}{M_{\rm eff}^2},
\ee
where $M_{\rm eff}$ is the effective mediator mass, and we define $L_{\rm eff}\equiv M_{\rm eff}^{-1}$ as the effective range of the Yukawa force. We can resolve the energy profile of a Fermi-ball again by the condition $dE_{\rm FB}/dR|_{R_{\rm FB}}=0$ after adding \Eq{EYuk_sim} into \Eq{EFB}. This is a cubic equation of $R^2$,
\be
R^4\frac{dE_{\rm FB}}{dR}=a(R^3)^2+b(R^2)^2+cR^2+d=0,
\ee
which can be transferred into the standard form
\be\label{u}
u^3+pu+q=0,
\ee
by $u=R^2+b/(3a)$, where
\be\begin{split}
p\approx&~-\frac{3}{16}\left(\frac{3}{2\pi}\right)^{2/3}\frac{Q_{\rm FB}^{4/3}}{U_0}\left(1-\frac{5M_i^2T^2}{108U_0}\right),\\
q\approx&~\frac{9y_\chi^2}{32\pi^2}\frac{L_{\rm eff}^2Q_{\rm FB}^2}{U_0}\left(1-\frac{M_i^2T^2}{12U_0}\right)+\frac{Q_{\rm FB}^2}{128U_0^2}\left(T^2+\frac{3M_i^2}{2\pi^2}\right).
\end{split}\ee
Now we can use the knowledge of the cubic equation to discuss the energy profile of a Fermi-ball and derive the collapse condition.

\begin{figure}
\centering
\subfigure{
\includegraphics[scale=0.295]{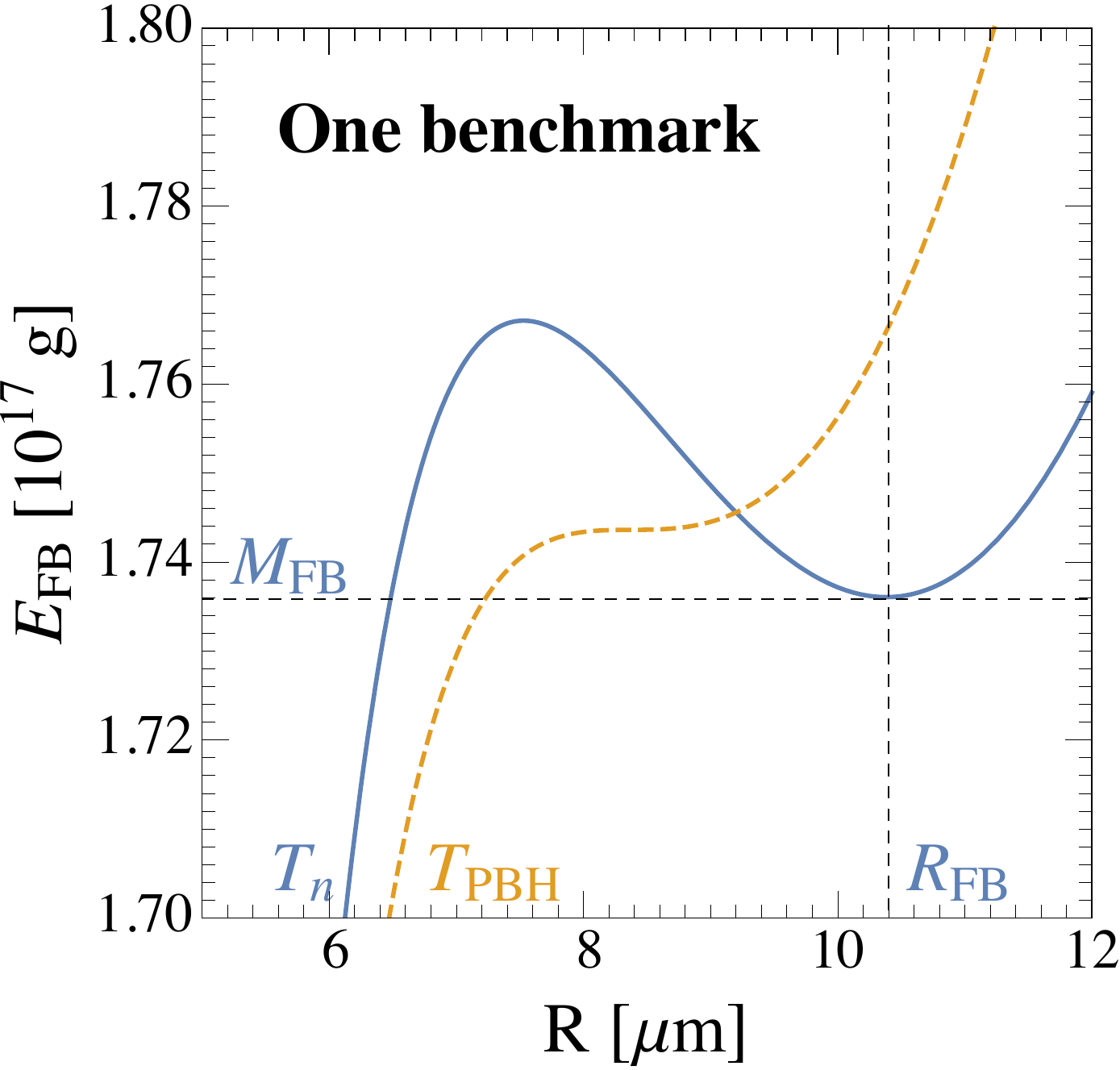}}
\subfigure{
\includegraphics[scale=0.295]{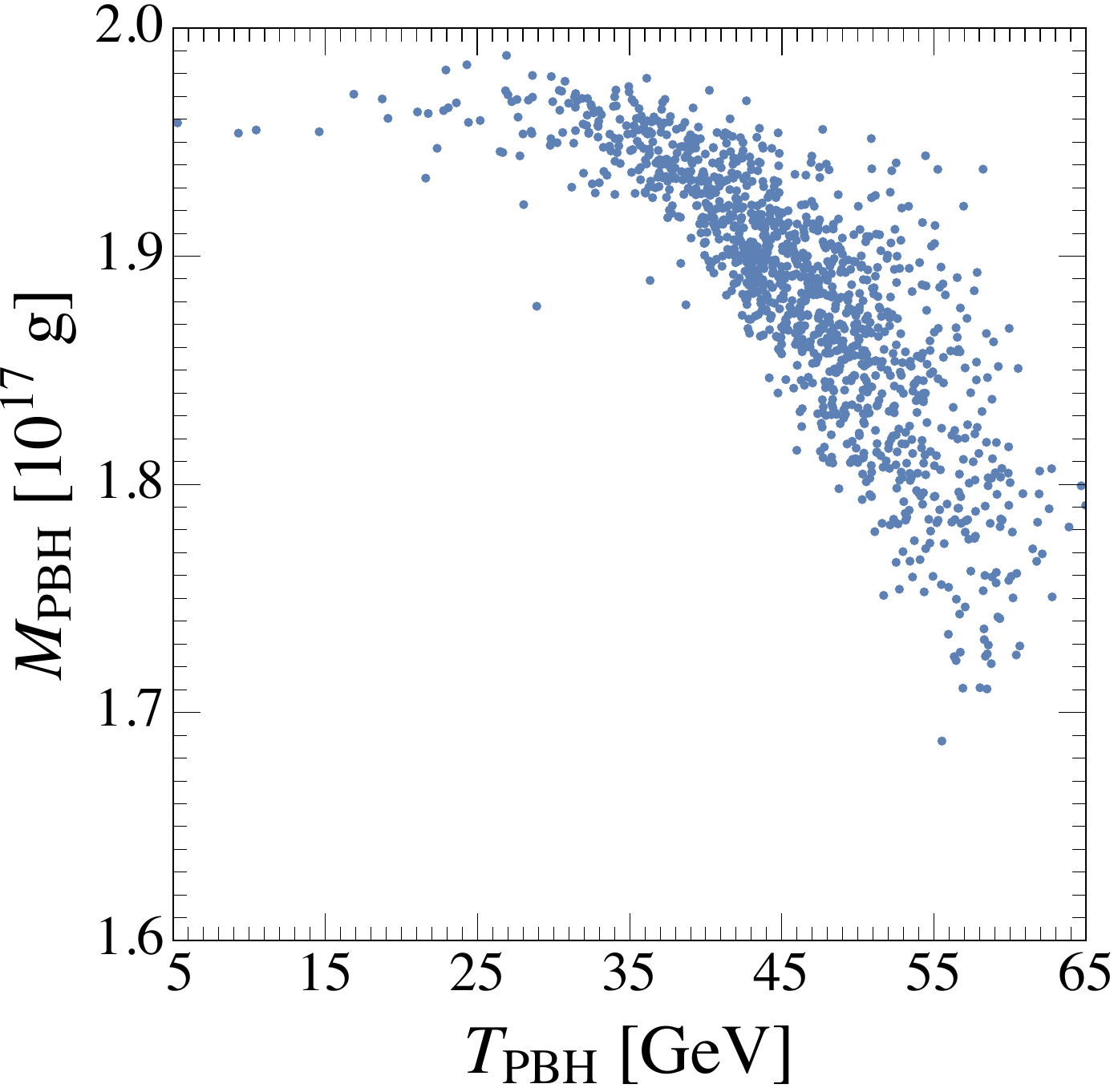}}
\caption{Left: illustration of the collapse of a Fermi-ball. At formation ($T_n$, blue curve), there is a local minimum for $E_{\rm FB}$, which can be explained as the Fermi-ball solution; however, when the Fermi-ball cools to $T_{\rm PBH}$ (dashed orange curve), $E_{\rm FB}$ is not bounded below, and the Fermi-ball collapses into a PBH. The benchmark: $\mu^2=(21.37~{\rm GeV})^2$, $\lambda=0.8253$, $a_1=-2338~{\rm GeV}$, $a_2=2.797$, $b_2=(839.1~{\rm GeV})^2$, $b_3=-66.32~{\rm GeV}$, $b_4=2.661$. Right: the masses and formation temperatures of the PBHs.}
\label{fig:collapse}
\end{figure}

Define the discriminant
\be\label{LphiU0}
\Delta=\left(\frac{q}{2}\right)^2+\left(\frac{p}{3}\right)^3,
\ee
which is a $T$-dependent value, as $M_i$, $U_0$ and $L_{\rm eff}$ are all functions of temperature. When $\Delta<0$, there are two positive roots for $u$ in \Eq{u}, with the larger one corresponding to the Fermi-ball radius $R_{\rm FB}$, and the smaller one giving the local maximum of the energy. When $T$ decreases, $\Delta$ increases. When $\Delta$ reaches 0, there is only one real root for $u$ and the energy profile is not bounded below any more. Physically, this is the collapse condition of a Fermi-ball to a PBH; written in terms of effective range of force, we obtain
\begin{widetext}
\begin{multline}
L_{\rm col}\approx\frac{1}{y_\chi}\left(\frac{2\pi}{3\sqrt3}\right)^{1/2}\left(\frac{2\pi}{3}\right)^{1/6}\frac{R_{\rm FB}}{Q_{\rm FB}^{1/3}}\left(1-\frac{(\sqrt3-1)M_i^2R_{\rm FB}^2}{3^{4/3}(2\pi)^{2/3}Q_{\rm FB}^{2/3}}\right)\\
-\frac{2\pi^2(\sqrt3-1)R_{\rm FB}^3T^2}{3^{13/4}y_\chi Q_{\rm FB}}\left(1+\frac{(52-30\sqrt{3})^{1/3}R_{\rm FB}^2M_i^2}{(9\pi Q_{\rm FB})^{2/3}}\right),
\end{multline}
\end{widetext}
and $L_{\rm eff}\geqslant L_{\rm col}$ means collapse. At $T_n$, $L_{\rm eff}<L_{\rm col}$, thus we have the Fermi-ball formation; while as the Fermi-ball cools, $L_{\rm eff}$ increases. When $L_{\rm eff}=L_{\rm col}$, a Fermi-ball collapses into a PBH. The collapse temperature is defined as $T_{\rm PBH}$, and the PBH inherits its mother Fermi-ball's mass. The collapse of a Fermi-ball is illustrated in the left panel of Fig.~\ref{fig:collapse} by a benchmark selected from the Fermi-ball data points derived in Section~\ref{sec:FB}.

Collapsing into a PBH is just one possible fate of a Fermi-ball. In general, when $T$ decreases and $L_{\rm eff}$ increases, there could be three different final states for a Fermi-ball:
\begin{enumerate}
\item $L_{\rm eff}$ reaches $L_{\rm col}$ at some temperature $T_{\rm PBH}$, and then a Fermi-ball collapses into a PBH.
\item $L_{\rm eff}<L_{\rm col}$ is always satisfied, even at $T=0$. The Fermi-balls survive today as a soliton DM candidate.
\item $L_{\rm eff}<L_{\rm col}$, but at some temperature $T_{\rm eva}$ the false vacuum disappears and there is only one minimum (the EW symmetry breaking vacuum) for the scalar potential. In that case, Fermi-balls will evaporate to free $\chi$'s at $T_{\rm eva}$.
\end{enumerate}
We find that among the Fermi-ball data points in Section~\ref{sec:FB}, around 72\% can collapse to PBHs, 1.1\% can survive until today to be soliton DM, and the others evaporate. Final state 3 is a novel feature from a two-field FOPT. Although Fermi-ball DM is also an interesting scenario~\cite{Hong:2020est,Marfatia:2021twj}, we will focus on the PBH data points hereafter. In the right panel of Fig.~\ref{fig:collapse}, we show the collection of mass and collapse temperature of the PBHs. One can see that the PBH masses are quite similar to the Fermi-ball masses in Fig.~\ref{fig:FB}, and typically $M_{\rm PBH}\sim\mO(10^{17}~{\rm g})$, lying in the allowed mass region for a 100\% DM contribution~\cite{Carr:2020gox,Carr:2020xqk,Green:2020jor}. This result is not surprising, because it is actually the motivation of choosing $\eta_\chi=10^{-8}$ as the benchmark in Section~\ref{sec:FB}. The PBH mass could be roughly estimated by~\cite{Kawana:2021tde}
\begin{multline}
M_{\rm PBH}\sim1.4\times10^{21}~{\rm g}\times v_b^3\left(\frac{\eta_\chi}{10^{-3}}\right)\times\\ 
\left(\frac{100}{g_*}\right)^{1/4}\left(\frac{100~{\rm GeV}}{T_*}\right)^2\left(\frac{100}{\beta/H}\right)^3\alpha^{1/4},
\end{multline}
thus $v_b$ and $\eta_\chi$ as well as the FOEWPT strength and duration affect the PBH mass, and we have tuned the parameters such that the resultant $M_{\rm PBH}$ lies in the allowed DM range.

So far, we have been working under the SM extended with two singlets: a real scalar $S$ and a Dirac fermion $\chi$. We have discussed the PBH formation in the FOEWPT, and evaluated the PBH mass profile. However, there are still two questions remaining unsolved,
\begin{enumerate}
\item How large is the relic abundance of the PBH?
\item What is the origin of the $\chi$-asymmetry $\eta_\chi$?
\end{enumerate}
As for the first question, since we already get $n_{\rm FB}^*$ (the number density of Fermi-balls at $T_n$, see the discussion below \Eq{R*}), the adiabatic expansion of the Universe would give the current density as $n_{\rm PBH}=n_{\rm FB}^*s_0/s(T_n)$, with $s_0\approx2891.2~{\rm cm}^{-3}$ the cosmic entropy density today~\cite{ParticleDataGroup:2020ssz}. By this we can obtain the relic density $\Omega_{\rm PBH}h^2=n_{\rm PBH}M_{\rm PBH}(8\pi h^2)/(3H_0^2M_{\rm Pl}^2)$, where $H_0$ is the current Hubble constant. As Ref.~\cite{Kawana:2021tde} already points out, the PBHs from Fermi-ball collapse tend to be overproduced, i.e. $\Omega_{\rm PBH}h^2\gtrsim\Omega_{\rm DM}h^2=0.12$. In fact, we obtain typically $\Omega_{\rm PBH}h^2\sim6000\times\Omega_{\rm DM}h^2$ for the data points in Fig.~\ref{fig:collapse}. Hence, an appropriate dilution mechanism must be combined to yield a PBH DM scenario. In next section, we build a concrete model to both generate the $\chi$-asymmetry and dilute the PBH density.

\section{Towards a complete model}\label{sec:UV}

\subsection{Generating the $\chi$-asymmetry}

We extend the model with a Dirac fermion $\psi$ and two real scalars $\phi_i$ with $i=1$, 2. All those particles are singlets under the SM gauge groups. The relevant Lagrangian reads
\be\label{UV}\begin{split}
\mL=&~\sum_{i=1}^2\left(\frac12\partial_\mu\phi_i\partial^\mu\phi_i-\frac12M_i^2\phi_i^2\right)\\
&~+\bar\psi\left(i\slashed{\partial}-M_\psi\right)\psi-\sum_{i=1}^2\left(\lambda_i\phi_i\bar\chi\psi+\hc\right).
\end{split}\ee
A mass hierarchy $M_2\gg M_1$ is imposed, and the lighter scalar $\phi_1$ can decay via $\phi_1\to\chi\bar\psi/\bar\chi\psi$. We also allow $\phi_1$ to decay to the SM particles, and the relevant interactions are not listed in \Eq{UV}. As we will see, when ${\rm Im}[(\lambda_1^*\lambda_2)^2]\neq0$, an $\eta_\chi$ can be generated. In addition, we assume $\psi$ only feebly couples to the SM particles, thus it never thermalizes. At late time, after the formation of PBHs, $\psi$ will dominate the energy of the Universe, leading to an early matter domination era. After that, $\psi$ decays to SM particles and reheats the Universe, diluting the PBH density to satisfy today's DM observation. This subsection focuses on the generation of $\eta_\chi$, while the next subsection will discuss the decay of $\psi$ and the dilution of PBH.

\begin{figure}
\centering
\includegraphics[scale=0.7]{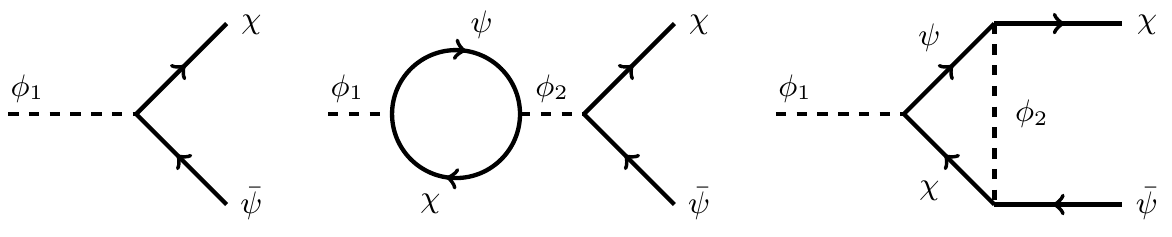}
\caption{The $\phi_1\to\chi\bar\psi$ process, Feynman diagrams related to the $\eta_\chi$ generation.}
\label{fig:asymmetry}
\end{figure}

The relevant Feynman diagrams for $\phi_1\to\chi\bar\psi$ are plotted in Fig.~\ref{fig:asymmetry}. For simplicity, we set $M_1\gg M_\chi$, $M_\psi$ so that the decay products can be treated as massless. At tree level, the decay width
\be
\Gamma_0(\phi_1\to\chi\bar\psi)=\frac{|\lambda_1|^2}{8\pi}M_1.
\ee
Similar calculation gives $\Gamma_0(\phi_1\to\bar\chi\psi)=\Gamma_0(\phi_1\to\chi\bar\psi)$. At one loop level, there are one self-energy diagram and one vertex correction diagram, and the standard loop calculation technique shows
\be\begin{split}
\epsilon_\chi=&~\frac{\Gamma(\phi_1\to\chi\bar\psi)-\Gamma(\phi_1\to\bar\chi\psi)}{\Gamma(\phi_1\to\chi\bar\psi)+\Gamma(\phi_1\to\bar\chi\psi)}\\
\approx&~-\frac{11}{48\pi}\frac{{\rm Im}[(\lambda_1^*\lambda_2)^2]}{|\lambda_1|^2}\frac{M_1^2}{M_2^2},
\end{split}\ee
where $M_1\ll M_2$ has been used. The $\chi$-asymmetry after reheating can be derived similar to the case of non-thermal leptogenesis~\cite{Asaka:1999yd}
\be
\eta_\chi=\frac{n_\chi-n_{\bar\chi}}{s}\approx\frac{3T_{\rm rh}}{2M_1}\epsilon_\chi{\rm Br}(\phi_1\to\chi\bar\psi/\bar\chi\psi),
\ee
where $T_{\rm rh}$ is the reheating temperature. At the same time a $\psi$-asymmetry $\eta_\psi=-\eta_\chi$ is generated.

\subsection{Evolution of $\psi$ and the dilution of PBH}\label{sec:psi}

After reheating, $\chi$ can thermalize due to the $-y_\chi S\bar\chi\chi$ interaction and the $|H|^2S^2$ portal couplings; however $\eta_\chi$ keeps unchanged because those interactions preserve the $U(1)_Q$ symmetry. On the other hand, we want $\psi$ to be out of equilibrium, i.e.
\be\label{no_washout}
n_\chi\ave{\sigma v_{\rm rel}}\ll H(T_{\rm rh})=2\pi\sqrt{\frac{\pi g_*}{45}}\frac{T_{\rm rh}^2}{M_{\rm Pl}},
\ee
where
\be
\ave{\sigma v_{\rm rel}}\approx\frac{|\lambda_1|^4}{64\pi^2}\frac{T_{\rm rh}^2}{M_1^4},
\ee
is the thermal average of the cross section of the annihilation $\chi\bar\psi\to\phi_1^*\to\bar\chi\psi$, and $n_\chi\approx 3\zeta_3T_{\rm rh}^3/(2\pi^2)$.

Next we consider the decay of $\psi$. Since it is not in the thermal bath, the yield
\be
Y_\psi+Y_{\bar\psi}=\frac{n_\psi+n_{\bar\psi}}{s}\approx\frac{3T_{\rm rh}}{2M_1}{\rm Br}(\phi_1\to\chi\bar\psi/\bar\chi\psi),
\ee
keeps as a constant. We assume the interaction $-\lambda_\psi\bar\ell_L\tilde H\psi$ such that $\psi$ can decay to SM leptons and the decay width
\be
\Gamma_\psi=\frac{|\lambda_\psi|^2}{16\pi}M_\psi.
\ee
$\psi$ will decay at $T_\psi$ which satisfies $\Gamma_\psi\approx H(T_\psi)$. If $\lambda_\psi$ is sufficiently small, $\psi$ can be treated as stable down to a very low temperature that it dominates the energy of the Universe. The domination of $\psi$ happens at $T_m$ when
\be
M_\psi Y_\psi\left(\frac{2\pi^2}{45}g_*T_m^3\right)\approx\frac{\pi^2}{30}g_*T_m^4.
\ee
If $T_\psi<T_m$, then below $T_m$ the Universe comes to a $\psi$ (matter) domination era. The decay of $\psi$ in such an era can produce extra entropy injection and dilute the PBH density. A more detailed discussion on an early matter era can be found in Refs.~\cite{Cosme:2020mck,Dutra:2021phm}, while here we adopt the simplified treatment that all $\psi$ decays quickly at $T_\psi$. This gives the entropy enhancement (dilution) factor~\cite{kolb1981early}
\be
\Delta_\psi=\frac{S_{\rm after}}{S_{\rm before}}\approx\left(\frac{T_\psi'}{T_\psi}\right)^3\approx 1.83\,\big\langle g_*^{1/3}\big\rangle^{3/4}\frac{M_\psi Y_\psi}{\sqrt{M_{\rm Pl}\Gamma_\psi}},
\ee
which can help to dilute the PBH density and realize the PBH DM scenario.

The PBH data points obtained in Section~\ref{sec:PBH} (see Fig.~\ref{fig:collapse}) typically require a dilution factor $\Delta_\psi\sim6000$ to give $\Omega_{\rm PBH}h^2/\Delta_\psi=0.12$. For the sake of this, we choose the relative phase between $\lambda_1$ and $\lambda_2$ to be $\varphi_\lambda=3\pi/4$ to enhance the CP violation which is $\propto\sin2\varphi_\lambda$. We further choose the following benchmark
\be\label{UV_bench}\begin{split}
&M_1\sim \frac{M_2}{350}\sim10^{10}~{\rm GeV},~M_\psi\sim1~{\rm TeV},\\
&|\lambda_1|\sim0.1,~|\lambda_2|\sim1,~|\lambda_\psi|\sim10^{-12},
\end{split}\ee
to rewrite the interaction rates and characteristic temperatures discussed above. First, the out-of-equilibrium condition of $\psi$, \Eq{no_washout}, can be normalized to
\begin{multline}
\frac{n_\chi\ave{\sigma v_{\rm rel}}}{H(T)}\Big|_{T_{\rm rh}}\approx1.36\times10^{-4}\times\left(\frac{|\lambda_1|}{0.1}\right)^4
\left(\frac{100}{g_*}\right)^{1/2}\\
\times\left(\frac{T_{\rm rh}}{4\times10^8~{\rm GeV}}\right)^3\left(\frac{10^{10}~{\rm GeV}}{M_1}\right)^4,
\end{multline}
and we can see that this is satisfied easily under the parameters we choose.

For the $\psi$-radiation equality,
\begin{multline}\label{Tm_new}
T_m=22.4~{\rm GeV}\times\left(\frac{1.0}{|\lambda_2|}\right)^2\left(\frac{10^{10}~{\rm GeV}}{M_1}\right)^2\\
\times\left(\frac{M_2}{3.5\times10^{12}~{\rm GeV}}\right)^2\left(\frac{M_\psi}{1~{\rm TeV}}\right)\left(\frac{\eta_\chi}{10^{-8}}\right),
\end{multline}
To avoid the complexity of an FOEWPT in a $\psi$ domination era, we require $T_m<T_n$. The $\psi$ decay temperature is
\be
T_\psi=3.82~{\rm MeV}\times\left(\frac{100}{g_*}\right)^{1/4}\left(\frac{\lambda_\psi}{10^{-12}}\right)\left(\frac{M_\psi}{1~{\rm TeV}}\right)^{1/2}.
\ee
Therefore, $T_\psi<T_m$ is satisfied. Finally, the dilution factor
\begin{multline}
\Delta_\psi=6.24\times10^3\times\left(\frac{1.0}{|\lambda_2|}\right)^2\left(\frac{g_*}{100}\right)^{1/4}\\
\times\left(\frac{M_2}{3.5\times10^{12}~{\rm GeV}}\right)^2
\left(\frac{10^{10}~{\rm GeV}}{M_1}\right)^2\\
\times\left(\frac{M_\psi}{10^3~{\rm GeV}}\right)^{1/2}\left(\frac{10^{-12}}{\lambda_\psi}\right)\left(\frac{\eta_\chi}{10^{-8}}\right),
\end{multline}
and hence the required $\Delta_\psi\sim6000$ can be realized. In other words, the parameter chosen in \Eq{UV_bench} can indeed provide a PBH DM scenario for our model. We have also checked that $T_\psi\Delta_\psi^{1/3}\gtrsim1$ MeV, thus the reheated temperature after $\psi$ decay is higher than $T_{\rm BBN}$, as required by the cosmological observations. 

The $\chi$-asymmetry after dilution is $\eta_\chi/\Delta_\psi\sim1.60\times10^{-12}$. Since $\eta_\chi/\Delta_\psi=-\eta_\psi/\Delta$, the $\psi$ decay will generate a lepton asymmetry, which is two orders of magnitude smaller than the observed baryon asymmetry $\eta_B^{\rm obs}\approx10^{-10}$~\cite{ParticleDataGroup:2020ssz}. As $T_\psi$ is much smaller than the decoupling temperature of the EW sphalerons, which is around 130 GeV~\cite{Burnier:2005hp}, the lepton asymmetry will not be converted to a baryon asymmetry. Note that $\eta_\chi/\Delta_\psi$ is unchanged when changing $\eta_\chi$. This is because $M_{\rm PBH}\propto\eta_\chi$, and hence increasing $\eta_\chi$ will also increase $\Delta_\psi$, leaving a fixed $\eta_\chi/\Delta_\psi\sim10^{-2}\eta_B^{\rm obs}$ determined by the DM abundance. This is a feature determined by an FOPT with $T_n\sim100$ GeV, and is already noticed in Ref.~\cite{Hong:2020est}.

While above discussions already demonstrate that for the complete model can provide the necessary $\Delta_\psi\sim6000$ dilution factor for the PBH data points, we still need to check two issues about the consistency of the treatment. First, in \Eq{Tm_new}, we have used $\eta_\chi$ to express ${\rm Br}(\phi_1\to\chi\bar\psi/\bar\chi\psi)$, thus it is necessary to see whether the branching ratio is smaller than 1. Using the benchmark in \Eq{UV_bench}, one obtains ${\rm Br}(\phi_1\to\chi\bar\psi/\bar\chi\psi)\sim28.0\%$, thus this is acceptable. The second issue is to make sure the dominant decay channel of $\psi$ is indeed to $\ell H$, as $\psi$ can also decay to $\chi$ plus two SM particles via an off-shell $\phi_1$. For the chosen benchmark, it turns out that the three-body decay width is typically $10^{-8}\times\Gamma(\psi\to\ell H)$ due to the huge suppression from $(M_\psi/M_1)^4$. Therefore, our complete model is self-consistent.

\section{Phenomenology: GWs and collider signals}\label{sec:pheno}

This section discusses the phenomenology of our model. First, as a mechanism accompanied by an FOEWPT, our scenario predicts phase transition GWs; second, the fermion $\chi$ can manifest itself via the multi-lepton/jet and displaced vertex signals at the LHC.

{\bf Phase transition GWs.} It is well known that an FOEWPT can generate stochastic GWs via bubble collision, sound waves and magneto-hydrodynamics turbulence in the plasma~\cite{Mazumdar:2018dfl}. Typically a transition happens at $T_n\sim100$ GeV yields GW signals at frequency $f\sim{\rm mHz}$ at current Universe~\cite{Grojean:2006bp}. As this frequency lies in the sensitive region of the near-future space-based detectors such as LISA~\cite{Audley:2017drz}, BBO~\cite{Crowder:2005nr}, TianQin~\cite{Luo:2015ght,Hu:2017yoc}, Taiji~\cite{Hu:2017mde,Guo:2018npi} and DECIGO~\cite{Kawamura:2011zz,Kawamura:2006up}, we expect the FOEWPT can be explored experimentally in the 2030s. The GWs in the singlet scalar extended SM is already extensively studied~\cite{Vaskonen:2016yiu,Alanne:2019bsm,Gould:2019qek}, and here we just focus on the GW signals from the PBH data points.

\begin{figure}
\centering
\subfigure{
\includegraphics[scale=0.285]{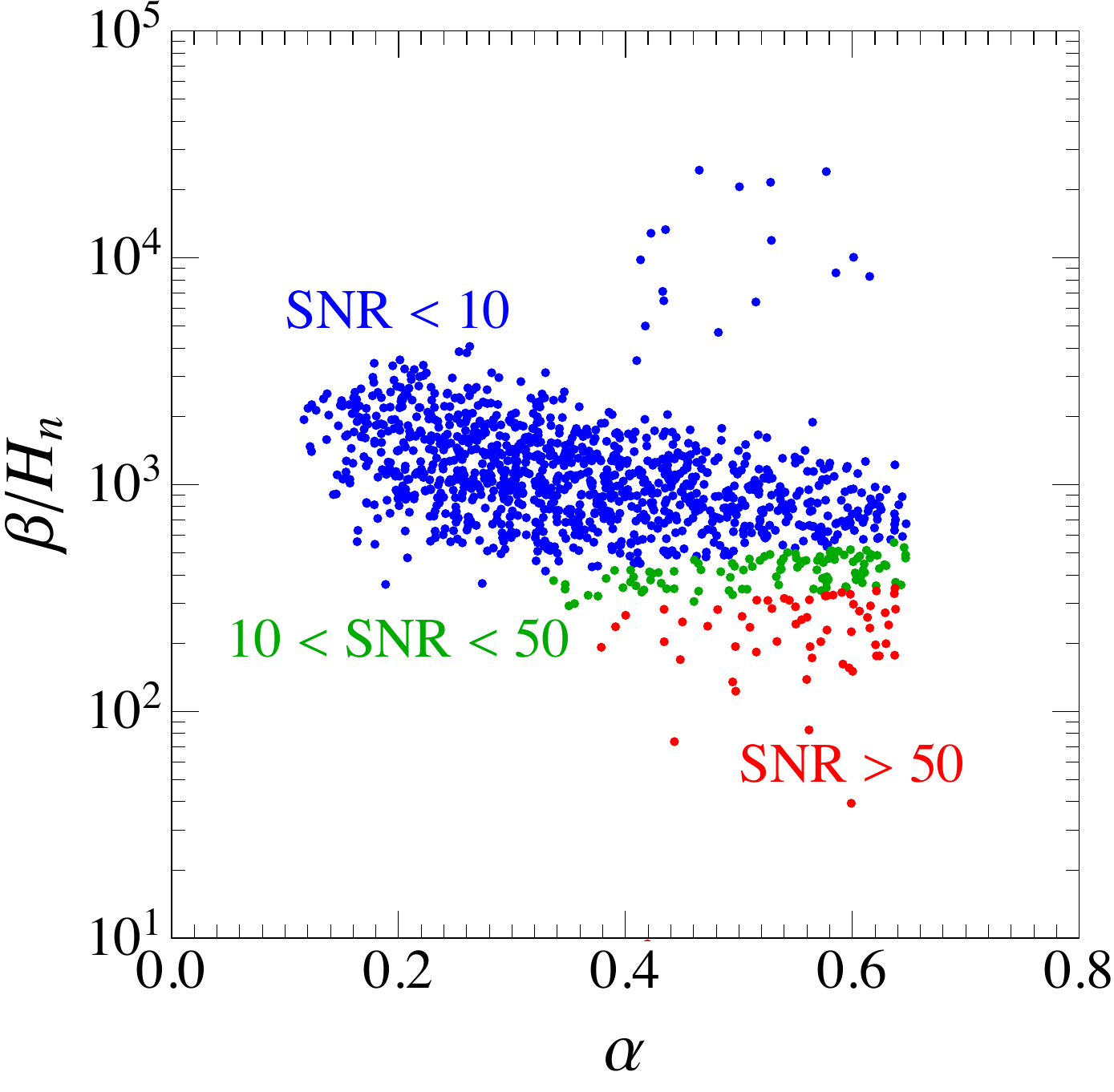}}
\subfigure{
\includegraphics[scale=0.235]{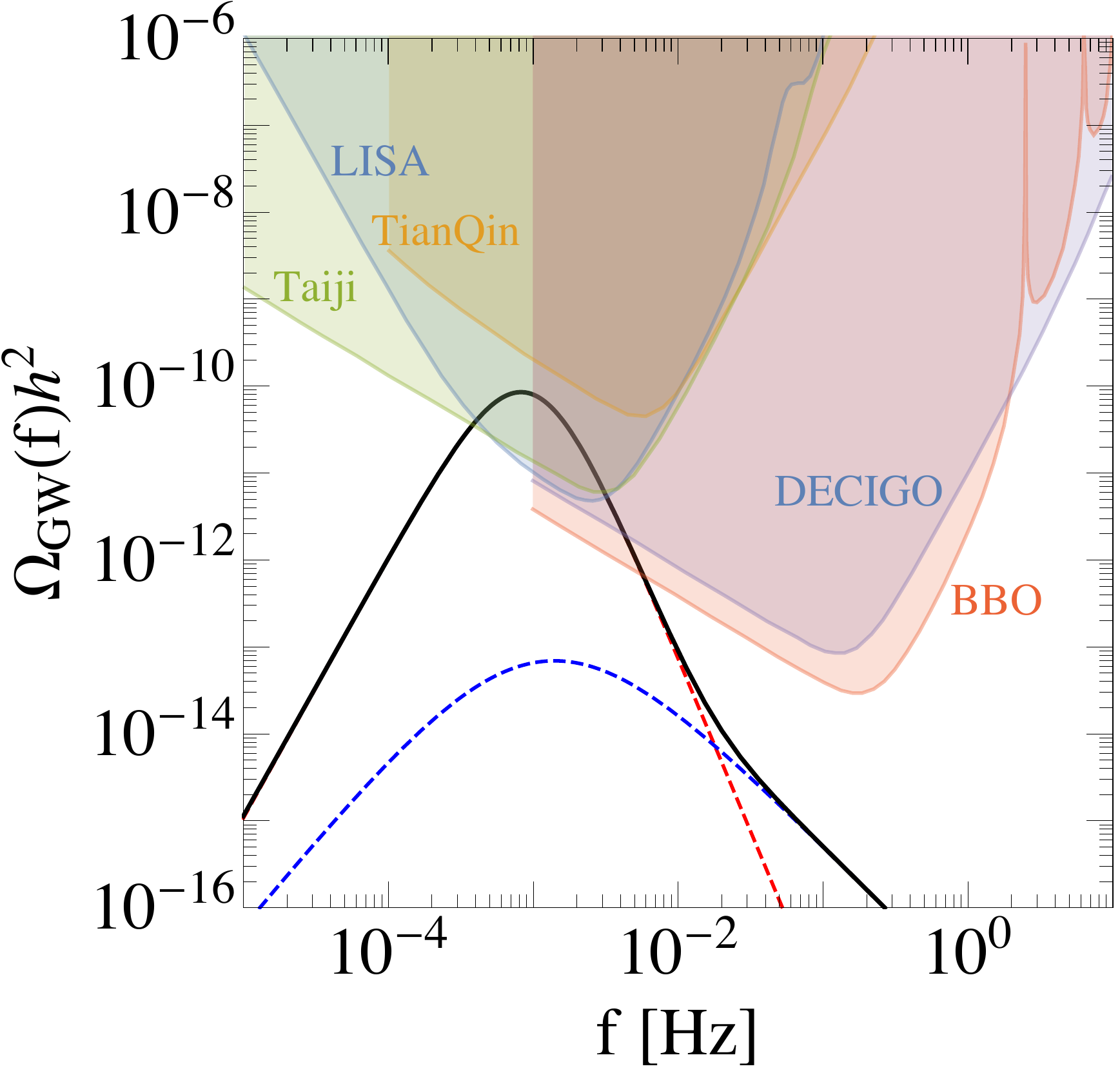}}
\caption{Left: collection of SNRs at the LISA detector for the PBH data points. Right: one GW spectrum benchmark, in which the black solid curve is the total signal, while the red and blue dashed curves are respectively the sound wave and turbulence contributions. The benchmark parameter is the same as the left panel of Fig.~\ref{fig:collapse}.}
\label{fig:GWs}
\end{figure}

The GW spectrum $\Omega_{\rm GW}(f)$ can be obtained as a function of the FOEWPT parameters $(\alpha, \beta/H_n, T_n,v_b)$~\cite{Grojean:2006bp,Caprini:2015zlo,Caprini:2019egz}. Given the spectrum, the signal-to-noise ratio (SNR)
\be
{\rm SNR}=\sqrt{\mT\int_{f_{\rm min}}^{f_{\rm max}} df\left(\frac{\Omega_{\rm GW}(f)}{\Omega_{\rm dec}(f)}\right)^2},
\ee
can be evaluated, where $\Omega_{\rm dec}$ is the sensitivity curve of the GW detector under consideration, and $\mT$ is the corresponding data-taking duration. For example, for the LISA detector, $\mT=9.46 \times 10^7$ s which is around four years~\cite{Caprini:2019egz}. For the detection threshold of SNR, we adopt ${\rm SNR}>10~(50)$ for the six-link (four-link) configuration LISA~\cite{Caprini:2015zlo}. The collection of SNRs of the PBH data points is shown in the left panel of Fig.~\ref{fig:GWs},\footnote{The suppression factor from the finite duration of sound wave period is included~\cite{Ellis:2018mja,Ellis:2019oqb,Guo:2020grp}.} and one benchmark of the GM spectrum is shown in the right panel of the same figure.

\begin{figure}
\centering
\subfigure{
\includegraphics[scale=0.295]{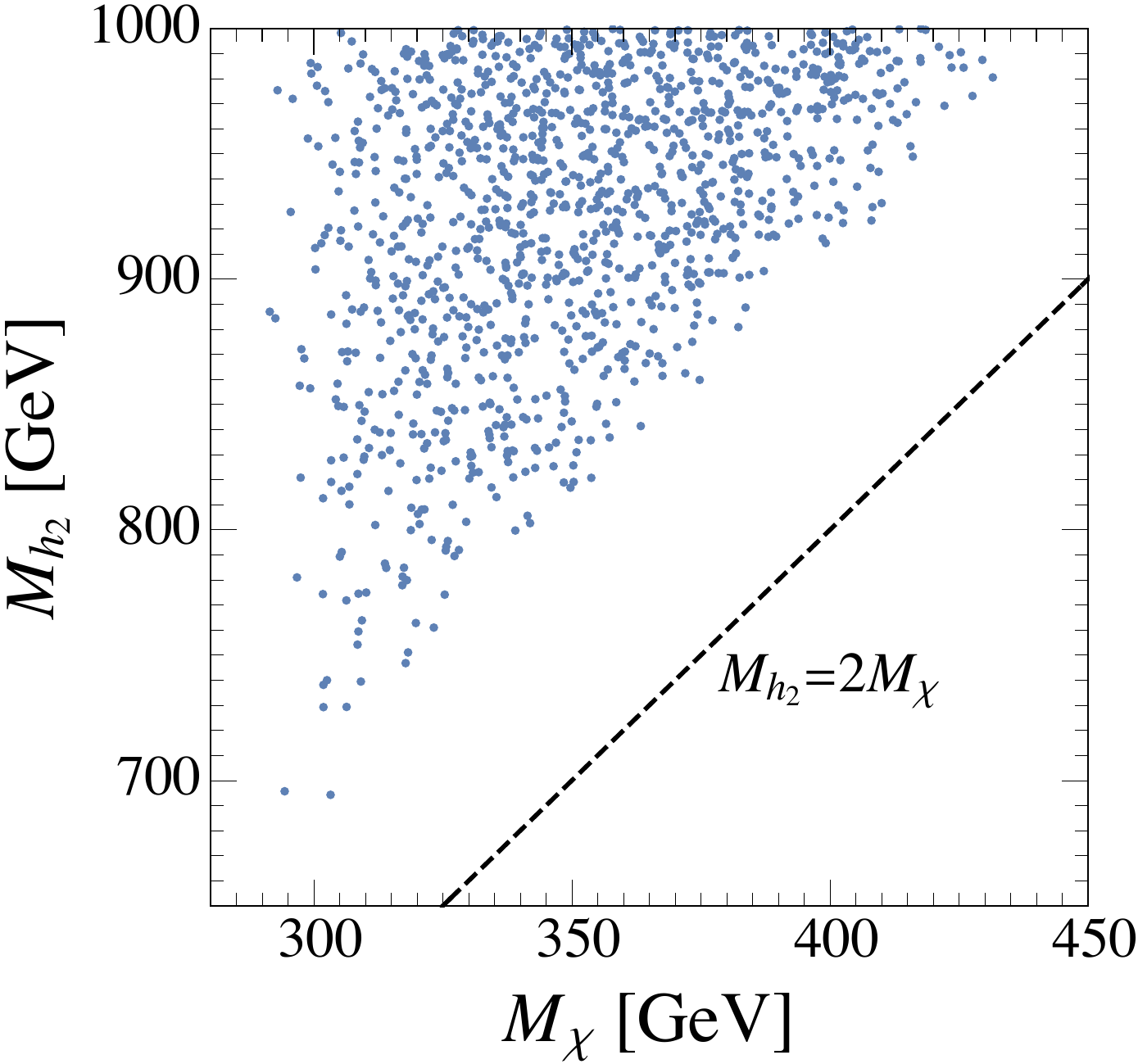}}
\subfigure{
\includegraphics[scale=0.28]{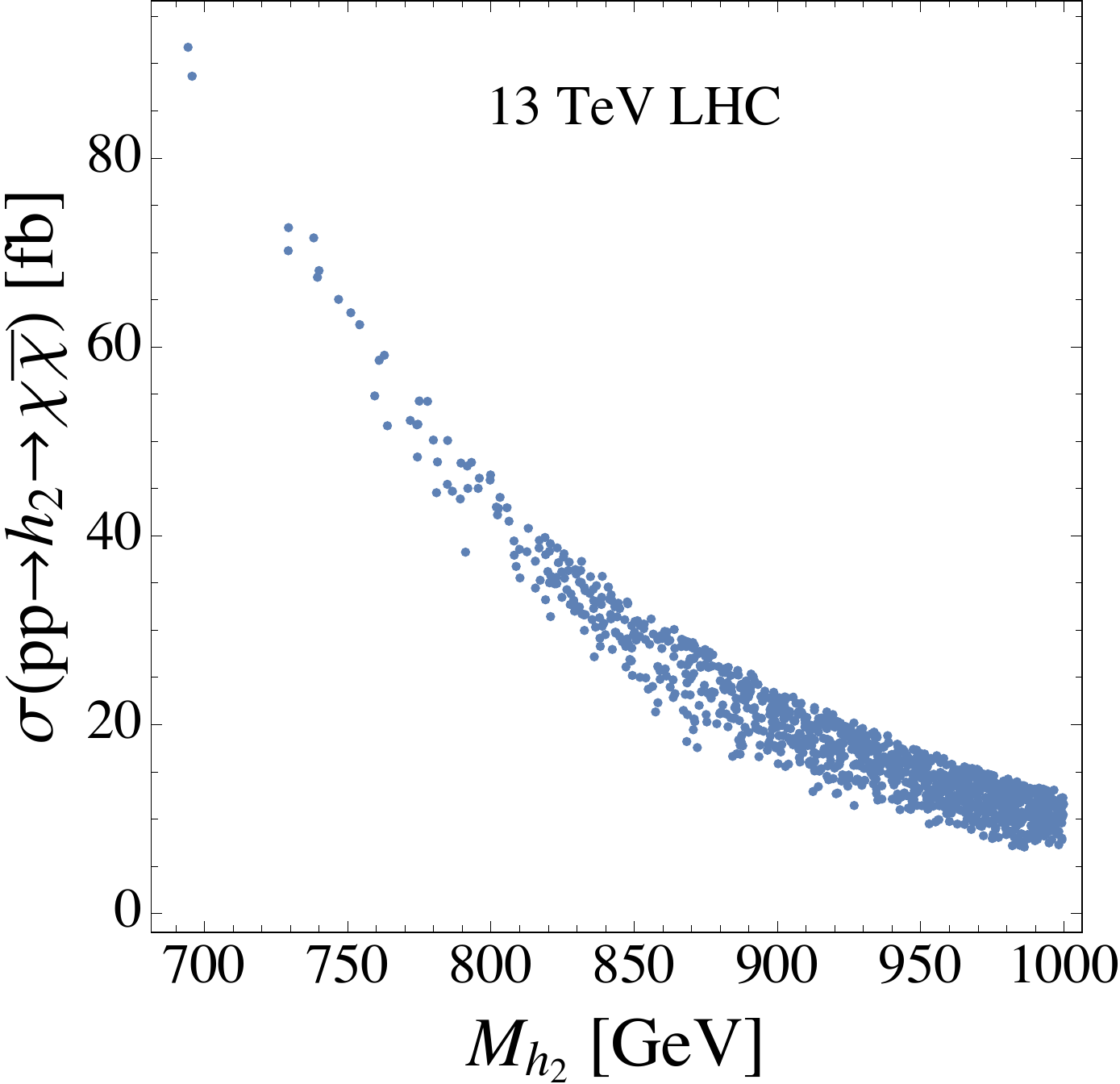}}
\caption{Left: the $M_{h_2}$ and $M_\chi$ distribution of the PBH data points. Right: collection of $pp\to h_2\chi\bar\chi$ cross sections at the 13 TeV LHC and the decay lengths of $\chi$.}
\label{fig:collider}
\end{figure}

{\bf Signals at the collider.} Current or future high energy colliders can probe the FOEWPT of the real scalar extended SM via the on-shell production of the singlet-like boson $h_2$, the deviation of Higgs couplings, etc~\cite{Profumo:2007wc,Profumo:2014opa,Alves:2018jsw,Huang:2018aja,Alves:2018oct,Alves:2019igs,Alves:2020bpi,Liu:2021jyc}. In this paper we are mainly interested in the collider signals of our model at the current LHC. Different from the real scalar extended SM, our model contains an additional fermion $\chi$. It turns out that $M_{h_2}>2M_\chi$ for the PBH data points, as shown in the left panel of Fig.~\ref{fig:collider}, where $M_\chi\equiv|M_0+y_\chi v_s|$ is the $\chi$ mass at true vacuum today. Due to the large $y_\chi\sim8$, $h_2$ decays dominantly to $\chi\bar\chi$, and the branching ratio is $\gtrsim85\%$. That means $pp\to h_2\to\chi\bar\chi$ is the main signal channel, in which $h_2$ is produced via gluon gluon fusion and the $h$-$S$ mixing.\footnote{We have verified that the $h_2\to h_1h_1$ and $h_2\to W^+W^-/ZZ$ channels are not reachable even at the HL-LHC, due to the low branching ratios.} The production rates are shown in the right panel of Fig.~\ref{fig:collider}.

To evade current direct detection searches for DM, the $\chi$ fermion in our scenario cannot be stable. The $\chi$ fermions in the true vacuum consist of two parts: the asymmetric component, from the penetration of $\chi$'s from the false vacuum; the symmetric component, from the $S\leftrightarrow\chi\bar\chi$ interaction. The latter component is negligible after freeze-out, since $y_\chi\sim8$ provides a large annihilation cross section $\sigma(\chi\bar\chi\to S)$. For the former component, we obtain $Y_\chi=(1-F_\chi^{\rm trap})\eta_\chi$, and the spin-independent scattering cross section on a nucleon can be evaluated using the formulae in Ref.~\cite{Li:2014wia} (the suppression factor from $\Omega_\chi/\Omega_{\rm DM}$ is taken into account, and $\Omega_\chi$ is the abundance after $\psi$ decay dilution). We found that for the PBH data points $\sigma_{\rm SI}\gtrsim5\times10^{-45}~{\rm cm}^2$, which is already excluded if $\chi$ can survive today~\cite{PandaX-4T:2021bab}. Therefore, we need to let $\chi$ decay.

The way to decay $\chi$ is similar to that is applied to $\psi$: we assume a vertex $\lambda_\chi\bar\ell_L\tilde H\chi$, which triggers the decay channel $\chi\to\ell^-W^+/\nu_\ell Z/\nu_\ell h$, and $\Gamma_\chi=|\lambda_\chi|^2M_\chi/(16\pi)$. But be aware this is the decay width in the true vacuum; in the false vacuum, $\chi$ is typically lighter than the scalar degrees of freedom, thus $\chi$ experiences a three-body decay to the SM light fermions via an off-shell scalar (including $h_1$, $h_2$ and the Goldstones). We use the {\tt MadGraph5\_aMC@NLO}~\cite{Alwall:2014hca} package to calculate the corresponding decay width, and require $\chi$ \emph{in the false vacuum} decays below $T_{\rm PBH}$ to ensure the Fermi-ball stability until collapse. A $\lambda_\chi\sim10^{-4}-10^{-3}$ can satisfy this condition. On the other hand, the $\chi$ fermions in \emph{the true vacuum} should decay above $T_{\rm BBN}\approx1$ MeV to evade the BBN constraint, yielding a rather weak bound that $\lambda_\chi\gtrsim10^{-13}$.

As shown above, the allowed $\lambda_\chi$ lies in a vast region, and so does the $\chi$ life time, which can be as short as $10^{-30}$ s or as long as 1 s. Therefore, the reaction $pp\to h_2\to\chi\bar\chi$, $\chi$ can provide different signals at the LHC. If life time $\tau_\chi\lesssim10^{-10}$~s, the subsequent decay $\chi\to\ell^-W^+/\nu_\ell Z/\nu_\ell h$ gives multi-lepton or multi-lepton plus jets signals, which might be probed by the supersymmetry searches such as Refs.~\cite{ATLAS:2021moa,ATLAS:2021yyr}. If $\chi$ is long-lived but the decay length $c\tau_\chi\lesssim10^3$ mm, $\chi$ might decay inside the LHC detectors, leaving the displaced vertex signals~\cite{Accomando:2016rpc,Deppisch:2018eth,Liu:2019ayx,Deppisch:2019kvs,Liu:2020vur}, and such searches are already performed by the ATLAS, CMS and LHCb collaborations~\cite{ATLAS:2019kpx,LHCb:2020akw,CMS:2021kdm,ATLAS:2020wjh}. Precision timing of the decay products can be also used to probe the long-lived $\chi$~\cite{Liu:2018wte}. For longer decay lengths, $\chi$ becomes missing transverse momentum, and we can only make use of the initial state radiation, probing $pp\to h_2j\to\chi\bar\chi j$ via the mono-jet signal~\cite{ATLAS:2021kxv}. In that case, the constraints are rather weak. We leave the detailed collider study and the possible correlation with the GW detection of our model for a future work.

\section{Conclusion}\label{sec:conclusion}

In this work, we propose an FOEWPT scenario that leads to PBH formation. The SM is extended with a singlet scalar $S$ to realize the FOEWPT, and with a singlet fermion $\chi$ to realize the trapping and Fermi-ball formation. As the Fermi-balls cool down, they collapse into PBHs due to the Yukawa attractive force. The model is further supplemented with two scalars $\phi_{1,2}$ and one fermion $\psi$ to generate the $\chi$-asymmetry and the necessary dilution factor for PBH density. We have demonstrated that the model can explain all DM via PBHs. This scenario can be tested via the GW signals at the future space-based interferometers and the multi-lepton/jet or displaced vertex searches at the current or future LHC.

Our model could be treated as a prototype of more general models. First, the FOEWPT in many models can be reduced to a ``Higgs plus singlet scalar'' pattern; second, a fermion that couples to the singlet is generally required by trapping. In addition, the necessary $\chi$-asymmetry and dilution factor are most easily realized via the decay of heavy particles. Therefore, we conclude that the model considered in this article has captured the most crucial and general features of the FOEWPT induced Fermi-ball and PBH formation mechanisms.

There are several directions to improve our work. As for the formation mechanism, we only considered $L_{\rm eff}<L_{\rm col}$ at $T_n$, i.e. during the FOEWPT the Yukawa force is negligible and hence the Fermi-balls can form. The formation of PBHs comes from a second-step collapse of the Fermi-balls. However, we also obtained parameter space with $L_{\rm eff}>L_{\rm col}$ at $T_n$, which means the false vacuum remnants collapse into PBHs without forming any solitons. The calculation of PBH profile in such a scenario requires the detailed treatment of the Yukawa interaction at the first stage, not like the treatment in Section~\ref{sec:PBH}, which just adds the Yukawa energy by hand to the existing Fermi-ball solution.

The model in this article can also be improved. For example, as shown in Section~\ref{sec:psi}, the the baryon asymmetry caused by $\chi$-asymmetry is negligible; a more elegant model may generate the $\chi$-asymmetry and baryon asymmetry simultaneously. In addition, the large Yukawa $y_\chi\sim8$ might cause the Landau pole problem at TeV scale. We remind the reader that first, the Yukawa coupling $y_\chi$ does not impact the FOEWPT dynamics, because the contribution from $\chi$ is Boltzmann suppressed. A large $y_\chi$ is only required to trap the fermions in the false vacuum. Second, a large $y_\chi$ is only necessary for the ``Higgs plus singlet scalar'' potential, because the VEV gap of $\ave{S}$ is fairly small during the FOEWPT; for other models, a smaller $y_\chi$ is possible. Finally, the large $y_\chi$ issue can be addressed by embedding the model into a strong dynamics framework and identifying $S$ as a pseudo-Nambu-Goldstone boson and $\chi$ as a composite lepton resonance.\footnote{See Refs.~\cite{Espinosa:2011eu,Bian:2019kmg,Xie:2020bkl,DeCurtis:2019rxl,Angelescu:2021pcd} for the FOEWPT in composite Higgs models.} In that case, the Landau pole can be relaxed to a higher scale because of the contributions from the TeV scale boson resonances, and at even higher scale, beyond the confinement of the strong dynamics, the $-y_\chi S\bar\chi\chi$ does not exist because the physical degrees of freedom change to more fundamental particles.

\acknowledgments

We thank Huai-Ke Guo, Chengcheng Han, Wei Liu and Mengchao Zhang for the very useful discussions. This work is supported by the National Science Foundation under grant number PHY-1820891, and PHY-2112680, University of Nebraska-Lincoln, and the University of Nebraska Foundation.

\appendix
\section{Trapping the fermions in old vacuum remnants}\label{app:trapping}

The trapping fraction can be derived by calculating the number of $\chi/\bar\chi$ fermions passing through the bubble wall. In the wall rest frame, the $\chi/\bar\chi$ distribution in the false vacuum is
\be
f_\chi^{\rm f.v.}(\p)=\frac{1}{e^{(\gamma_+\epsilon+\gamma_+v_+p_z)/T_*}+1},
\ee
where $\epsilon=\sqrt{M_i^2+|\p|^2}$, and the wall is taken as the $Oxy$ plane, and the $z<0$ region is the true vacuum. $\gamma_+$ is the Lorentz factor $(1-v_+^2)^{-1/2}$. The number density in the false vacuum is
\be
n_\chi^{\rm f.v.}=2\int\frac{d^3p}{(2\pi)^3}f_\chi^{\rm f.v.}(\p).
\ee

A $\chi$ fermion with $z$-component momentum $|p_z|>\sqrt{M_f^2-M_i^2}$ can pass through the wall. The particle current per unit area and unit time is then
\be
J_\chi^{\rm w.}=2\int\frac{d^3\p}{(2\pi)^3}\frac{-p_z}{|\p|}f_\chi^{\rm f.v.}(\p)\Theta\left(-p_z-\sqrt{M_f^2-M_i^2}\right),
\ee
where $\Theta$ is the Heaviside step function. The particle current can be transformed into the plasma frame by multiplying a time dilation factor $J_\chi=J_\chi^{\rm w.}/\gamma_b$. Therefore, in the true vacuum, the $\chi$ density caused by the penetration is $n_\chi^{\rm pene.}=J_\chi/v_b$. The fraction of trapped $\chi$ in false vacuum is defined as
\be
F_\chi^{\rm trap}=1-\frac{n_\chi^{\rm pene.}}{n_\chi^{\rm f.v.}}=J_\chi^{\rm w.}\frac{\sqrt{1-v_b^2}}{v_b}.
\ee
Note that the derivation of $F_\chi^{\rm trap}$ is valid only when $n_\chi^{\rm pene.}\ll n_\chi^{\rm f.v.}$, so that the $\chi/\bar\chi$ fermions in the false vacuum can be approximated as in equilibrium. Increasing the bubble velocity generally decreases the trapping rate because the $\chi$'s are more energetic in the wall frame that they are easier to penetrate into the bubble.

\bibliographystyle{apsrev}
\bibliography{references}

\end{document}